# Discrete Wavelet Transform-Based Prediction of Stock Index: A Study on National Stock Exchange Fifty Index

# Discrete Wavelet Transform-Based Prediction of Stock Index: A Study on National Stock Exchange Fifty Index

Dhanya Jothimani, Ravi Shankar, Surendra S. Yadav
Department of Management Studies,
Indian Institute of Technology Delhi, India
dhanyajothimani@gmail.com, ravi1@dms.iitd.ac.in, ssyadav@dms.iitd.ac.in

## Abstract

Financial time series such as stock price and exchange rates are, often, non-linear and non-stationary. Previously, many researchers have attempted to forecast those using statistical models and machine learning models. Statistical models assume the time series to be stationary and linear, thus resulting in large statistical errors. Though machine learning models, namely, Artificial Neural Networks (ANN) and Support Vector Regression (SVR) can model non-linear data quite effectively, they suffer from the problem of overfitting and are sensitive to parameter selection. Use of decomposition models has been found to improve the accuracy of predictive models. The paper proposes a hybrid approach integrating the advantages of both decomposition model (namely, Maximal Overlap Discrete Wavelet Transform (MODWT)) and machine learning models (ANN and SVR) to predict the National Stock Exchange fifty index. In first phase, the data is decomposed into a smaller number of subseries using MODWT. In next phase, each subseries is predicted using machine learning models (i.e., ANN and SVR). The predicted subseries are aggregated to obtain the final forecasts. In last stage, the effectiveness of the proposed approach is evaluated using error measures and statistical test. The proposed methods (MODWT-ANN and MODWT-SVR) are compared with ANN and SVR models. Further, it was observed that the return on investment obtained based on trading rules using predicted values of MODWT-SVR model is higher than that of Buy-and-hold strategy.

---

Corresponding Author: Dhanya Jothimani, dhanyajothimani@gmail.com

## 1. Introduction

Stock market can be viewed as a complex system that takes in a lot of information ranging from fundamental information to socio-political events and news to investors' behavior and produces an output (Novak & Velušček, 2015[1]). Due to complex market dynamics, there is an ongoing debate on "can stock prices be predicted accurately?". Finance literature suggests two approaches for predicting the movements of stock price: Fundamental analysis and Technical analysis. In fundamental analysis, the stock prices are predicted using fundamental indicators of a company such as Return on Equity (ROE), Earnings Per Share (EPS) and Price to Earnings (PE) ratio.

In technical analysis, the movement of stock price is predicted based on the behavior of previous stock price values. Technical analysts focus on the market timings. They use various tools such as charts, technical indicators and models to monitor the trend of price and volume over a long time period. The focus of technical analyst is to develop a model that is capable of predicting the stock price dynamics accurately.

According to widely accepted Random Walk theory, the markets move in a random and unpredictable manner. It reflects the efficient market hypothesis and states that if the publicly available information is reflected on the stock price then tomorrow's price is independent of today's price but the effect of the available information (Fama, 1970[2]).

Few studies showed the evidence of predictability of stock prices. Lo & MacKinlay (1988)[3], Chen (1991)[4], and Bilson, Brailsford, & Hooper (2001)[5] showed that macroeconomic factors aided in forecasting of stock prices. Yao, Tan, & Poh (1999)[6] used technical indicators (Relative Strength Indicator, Momentum and Moving Average) and lagged price index for predicting the movement of stock price. Later, combination of both fundamental and technical indicators was studied for stock index forecasting (Bettman, Sault, & Schultz, 2009)[7].

Various models have been adopted to predict the financial time series, especially stock price. Atsalakis & Valavanis (2013)[8] and Atsalakis & Valavanis (2009)[9] surveyed various traditional models and soft computing techniques for prediction of stock price, respectively.

Statistical models such as AutoRegressive (AR) models, Moving Average (MA), Autoregressive Moving Average (ARMA) and AutoRegressive Integrated Moving Average (ARIMA) assume that stock prices follow normal distribution and are stationary and linear. However, stock prices are non-linear and non-stationary. Generalized Autoregressive Conditional Heteroskedastic (GARCH), proposed by Bollerslev (1986)[10], and its extensions try to model the heterogeneity of volatility and handle irregularities. However, they fail to completely capture highly irregular phenomena in financial market (Matei, 2009)[11].

The ability of various machine learning techniques such as Artificial Neural Networks (ANN) and Support Vector Regression (SVR) to capture and model non-stationary and non-linear data has led to their wide applicability in time series forecasting. But these models are not without their limitations. They suffer from the problem of overfitting and getting trapped in local optima.

Accuracy of the forecasts can be improved by (1) improving the algorithms, (2) preprocessing the data, or (3) both. In data preprocessing, the data is transformed into a format that reveals certain characteristics. Decomposition of time series is one such technique, where the time series is deconstructed into several components. There are two types of decomposition models: (i) classical decomposition, and (ii) non-classical decomposition models. The most commonly used classical model of decomposing the time series into trend, seasonal and random components, works best with the linear time series. It ignores the random component and leads to a loss of information, thus, affecting the forecast accuracy (Theodosiou, 2011)[12].

Recently, several signal processing techniques like Maximal Overlap Discrete Wavelet Transform (MODWT) and Empirical Mode Decomposition (EMD) have been used for decomposing the series in time-frequency domain and time domain, respectively (Liu, Chen, Tian, & Li, 2012[13]; Lahmiri, 2014[14]). They fall under the category of non-classical decomposition model.

The paper presents an approach depicting two models, namely, hybrid MODWT-ANN and hybrid MODWT-SVR models to predict 1-step ahead forecasts for weekly National Stock Exchange Fifty price index, where the time series is first decomposed to different sub-series using MODWT. Then, these sub-series are predicted independently using two machine learning models and are aggregated to obtain the final forecasts. The hybrid MODWT-SVR model integrates the benefits of both MODWT and SVR. Similarly, the hybrid MODWT-ANN model highlights the advantages of both MODWT and ANN.

Apart from improvement in prediction accuracy of stock price index, another important area which interests the investors is the optimal timing for buying and selling the security to minimize loss or maximize returns. In this paper, few trading rules based on the predictive model are illustrated to guide the investors to make buying and selling decisions.

The flow of the paper is as follows: Section 2 explains the theory of the methodology adopted in this paper. The analysis of National Stock Exchange Fifty index is presented in Section 3. The results and the trading decisions are discussed in Section 4 followed by Conclusion in Section 5.

## 2. Research Methodology

### 2.1 Discrete Wavelet Transform (DWT)

Wavelet analysis helps to analyse localized variations of signal within a time series. Both the dominant modes of variability and their variations in time can be captured by decomposing a time series into time-scale (or time-frequency) space. Discrete Wavelet Transform (DWT) can decompose the signal in both time and frequency domain simultaneously. On the other hand, Fourier Transform decomposes the signal only in frequency domain; information related to occurrence of frequency is not captured and it eliminates the time resolution (Ortega & Khashanah, 2014[15]).

Any function *y(t)* can be decomposed by a sequence of projections onto wavelet basis:

$$s_{J,k} = \int y(t)\Phi_{J,k}(t)dt \tag{1}$$

$$d_{j,k} = \int y(t)\Psi_{j,k}(t)dt \tag{2}$$

where $J$ represents the number of multiresolution; $\Phi$ is the father wavelet and $\psi$ is the mother wavelet. $s_{J,k}$ and $d_{j,k}$ represent the smooth and the detailed coefficients, respectively. $\Phi_{j,k}$ and $\psi_{j,k}$ are scaling and translation of $\Phi$ and $\psi$, defined by

$$\Phi_{j,k}(t) = 2^{-j/2}\Phi(2^{-j}t - k) = 2^{-j/2}\Phi(\frac{t - 2^{j}k}{2^{j}}) \qquad (3)$$

$$\Psi_{j,k}(t) = 2^{-j/2}\Psi(2^{-j}t - k) = 2^{-j/2}\Psi(\frac{t - 2^{j}k}{2^{j}}) \qquad (4)$$

$s_{j,k}$ and $d_{j,k}$ are signal representation in wavelet domain.

The father wavelet and the mother wavelet approximate the smooth (low frequency) components and the detail (high frequency) components of the signal, respectively.

The decomposition of signal using DWT is shown in Figure 1. DWT operates on the principle of convolution. The wavelet at a scale acts as band-pass filter (band corresponds to scale). On decomposing the original signal $y(t)$, two components $A_1$ and $D_1$ are produced by convulating the signal with a decomposition low pass filter (*D_LP*) and a decomposition high pass filter (*D_H*), respectively. $A_1$ is decomposed into $D_2$ and $A_2$ by a larger scale and so on. $A_1$ is the approximation signal which shows the general trend of low frequency component. $D_1$ is detail of signal and is associated with high frequency component of the signal. The original series in terms of components can be represented as:

$$y(t) = A_J + \sum_{i=1}^{J} D_J \qquad (5)$$

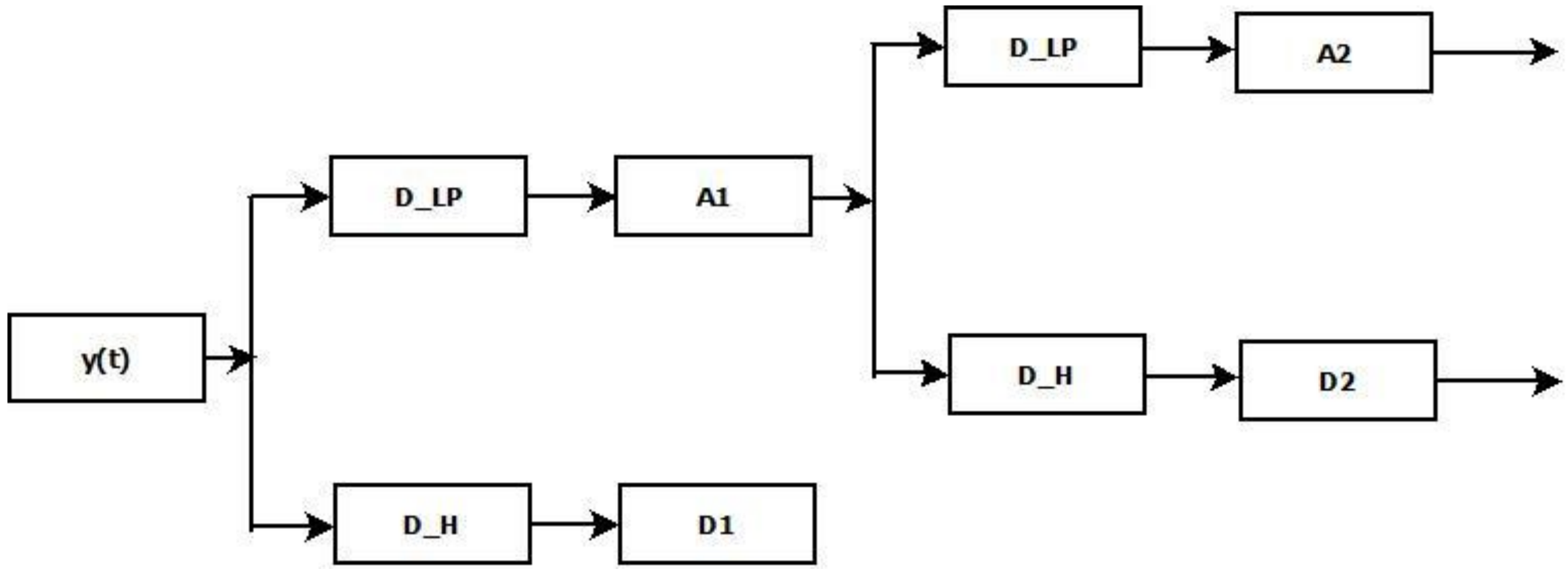

**Figure 1: Decomposition of Signal using DWT**

Various kinds of wavelets are available, for instance, Haar, Daubechies, Morlet and Mexican Hat. Morlet and Daubechies wavelets have applications in image processing and are subject to aliasing problem. Mexican Hat wavelets are expensive to compute. Haar wavelets are advantageous for time series analysis since they are capable of capturing fluctuations between adjacent observations (Ortega & Khashanah, 2014[15]; Lahmiri, 2014[14]; Murtagh, Starck, & Renaud, 2004[16]; Li, Li, Zhu, & Ogihara, 2002[17]).

One of the limitations of DWT is the requirement of length of the dataset to be dyadic (i.e., power of 2). Secondly, the output generated by DWT is highly dependent on origin of the signal being analyzed. A small shift in origin affects the outputs generated and this problem is called Circular shift. Due to circular shift, it is difficult to align the transformed signals with time. To overcome the above two limitations, a modification of DWT called Maximal Overlap Discrete Wavelet Transform (MODWT) is used. MODWT is circular shift invariant and is not limited by the dyadic length constraint; hence the signals are easier to interpret for time series analysis. DWT and MODWT are used interchangeably this point forward.

Practical applications of wavelets in Finance and Economics can be found in Gençay, Selçuk, & Whitcher (2002)[18], Ortega & Khashanah (2014)[15] and Lahmiri (2014)[14].

### 2.2 Artificial Neural Networks (ANN)

Inspired by biological neural networks (brain), Artificial Neural Networks (ANN) are family of models in machine learning that have been developed, and are used for function approximation, classification and regression. ANN models consist of densely interconnected computational units called neurons. The neurons are organized in three layers: input layer, hidden layer and output layer (See Figure 2). The connection between neurons of consecutive layers is weight (also called connection weight). The first layer in the network, input layer corresponds to input variables. The second layer in the network is hidden layer, which is connected to both input and output layer. The neurons in the hidden layer perform two operations: (i) they receive information forwarded from the neurons in the input layer; and (ii) multiply the information by a weight factor. Weighted information of all incoming connections is added in each neuron of the hidden layer and finally a bias term is included. The summed input and bias are processed using a non-linear function (sigmoid or hyperbolic tangent function) to obtain an output value which is then forwarded to the neurons in the output layer, the third layer of the network. The mathematical relationship between input *x(t)* and output *y(t)* is represented as

$$y(t) = w(0) + \sum_{j=1}^{u} w(j).f(w(0,j) + \sum_{i=1}^{v} w(i,j).x(t)) \quad (6)$$

where, *w(i,j) (i=0,1,2,...,p; j=1,2,...,u)* and *w(j) (j=0,1,2,...,u)* are the connection weights, *v* is the number of input nodes, *u* is the number of hidden nodes, and *f* is the activation function.

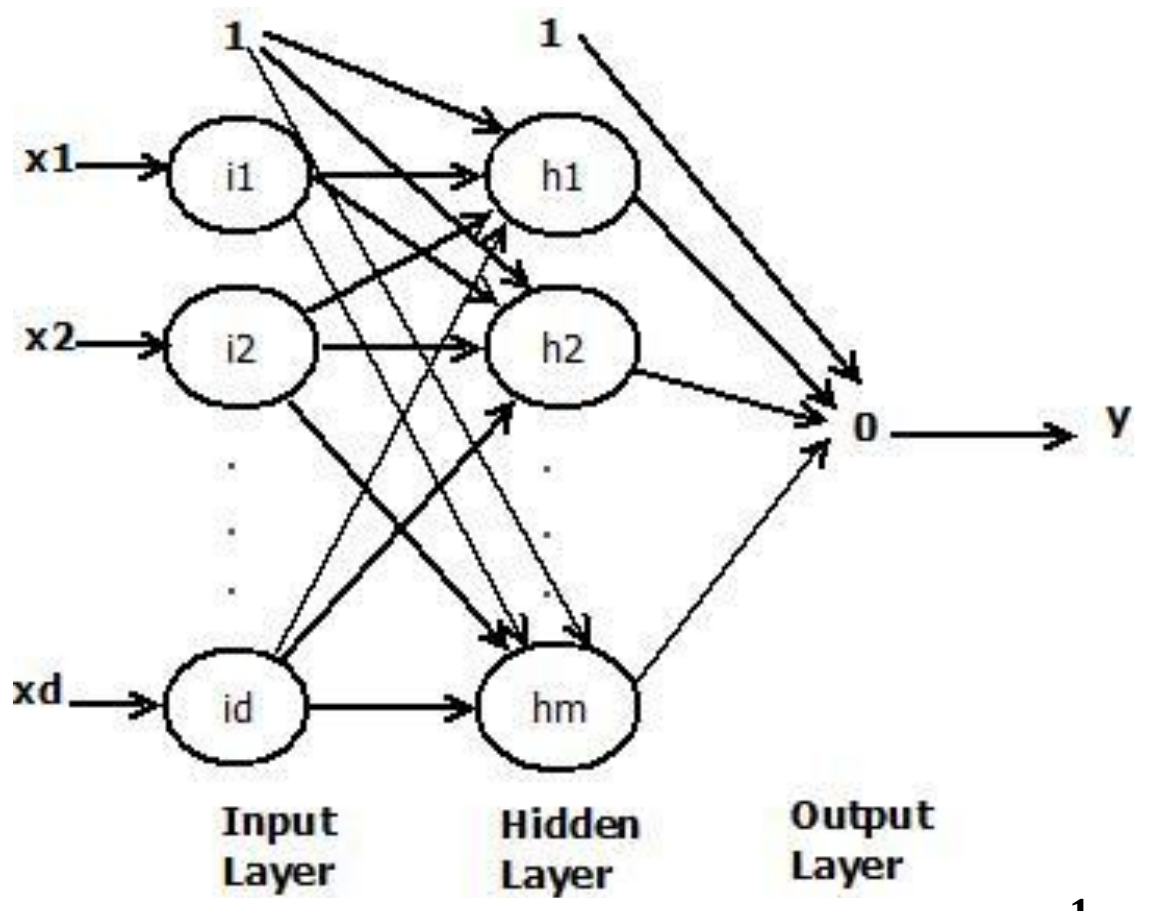

**Figure 2: Architecture of ANN model**[1]

The weights are adjusted so that a network can 'learn' to reproduce the relationship between output and input patterns. This phase is called training. This type of model is generally known as supervised model since both input and output data are provided during training phase. The output produced by the network can be compared with desired (original) output.

Most commonly used training algorithm is Backpropoagation (BP) whose objective function is to minimize the sum of squares of difference between the desired output $y_d(t)$ and network predicted output $y_p(t)$, i.e., error which is represented as

$$E = \frac{1}{2}\sum_{t=1}(y_d(t) - y_p(t))^2 \quad (7)$$

During training by BP algorithm, the error component is transmitted backwards from output layer to each neuron in the hidden layer (Figure 3). The weights and biases related to each neuron are adjusted based on the error component. The iterative process helps the network to converge and the process is carried out till the error ($E$) gets minimized. BP uses gradient descent

[1] $i_1,\ldots, i_d$ & $h_1,\ldots,h_m$ represent the neurons in the input layer and the hidden layer, respectively, O represents the output neurons and y is the obtained output (here it refers to the predicted values)

algorithm and is sensitive to parameters such as learning rate and momentum, which might override the possibility of finding an optimal solution. To overcome this limitation, Resilient Propagation (RP) proposed by Riedmiller & Braun (1993)[19] is used. Another advantage of RP is that it quickens the training process and helps to achieve superior performance (Liu et al., 2012[13]).

Once the network is trained, it can be used to predict for new unknown input values based on the knowledge obtained during the training phase. This is called testing phase. The training of a neural network model using BP algorithm is depicted in Figure 3.

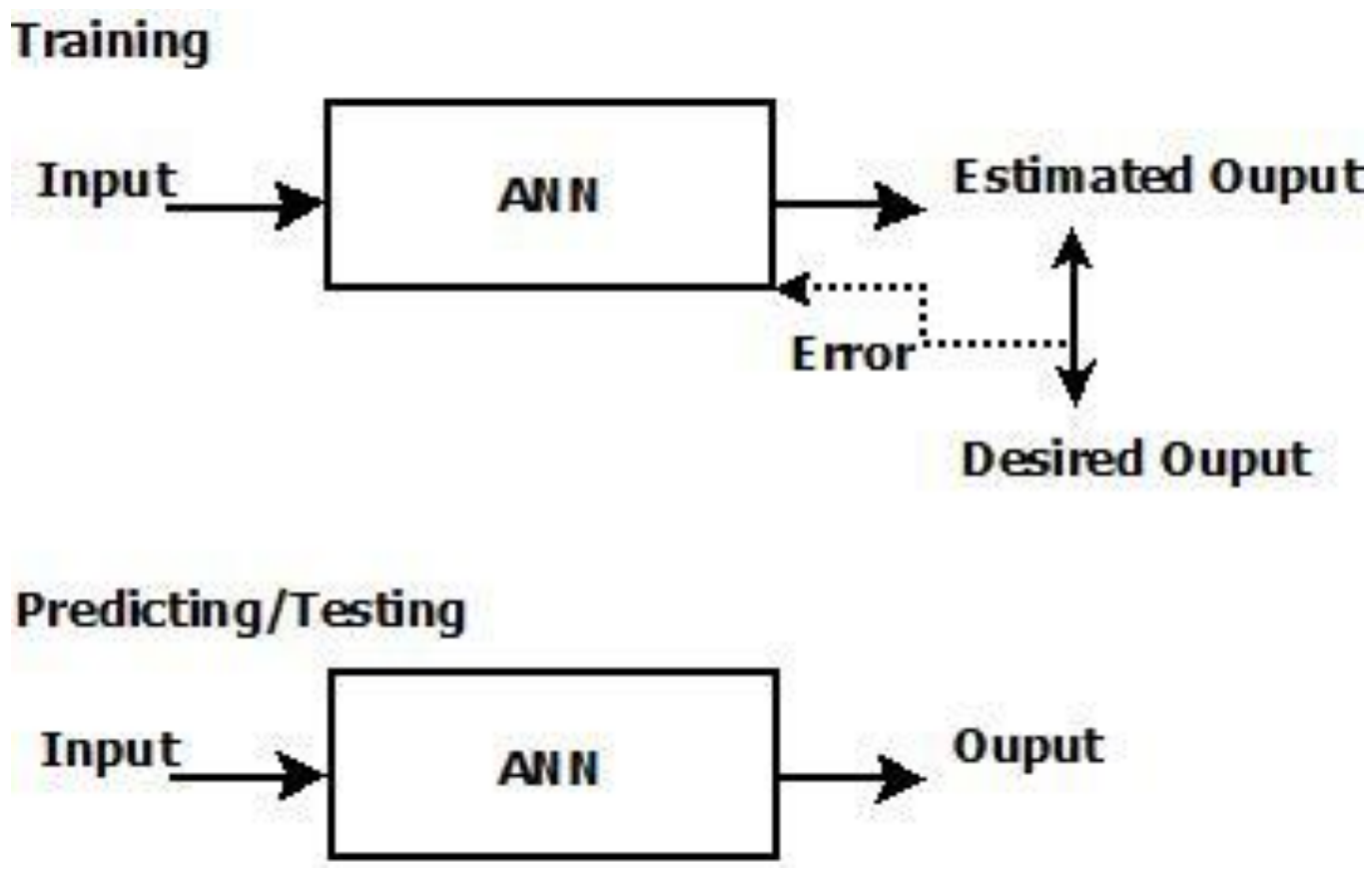

**Figure 3: Training and Testing of a ANN model using BP algorithm**

The reasons for wide applicability of ANN in time series forecasting are enumerated below (Al-Hnaity & Abbod, 2015[20]):

1. Unlike statistical models, ANN models require few prior assumptions. They are generally data-driven and adaptive.
2. ANN models are capable of modelling and predicting the non-linear data.

3. ANN models can approximate any continuous function with satisfactory accuracy.

## 2.3 Support Vector Regression (SVR)

Based on VC theory (VC-Vapnik, Chervonenkis) or statistical learning theory, Support Vector Regression (SVR) has gained popularity in the recent years for modeling non-linear regression. Similar to ANN, SVR is a data-driven supervised learning model. The schematic diagram of SVR is shown in Figure 4.

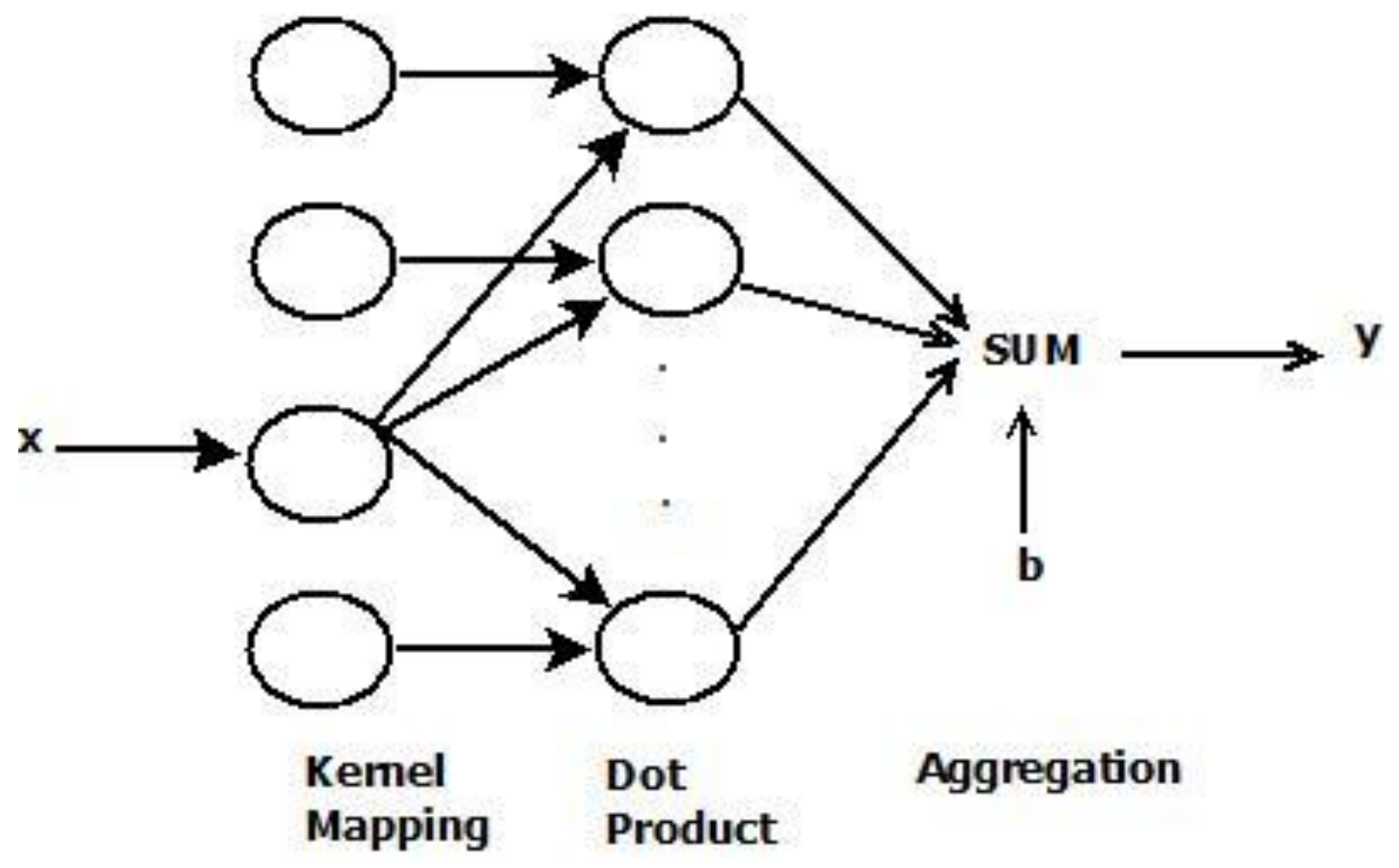

**Figure 4: Schematic diagram of SVR**

The SVR model (Vapnik, 1995[21]) is represented as:

$$f(x) = (\mathbf{z}.\phi(x)) + b \tag{8}$$

where $b$ is bias, **z** is a weight vector, and $\Phi(x)$ is a kernel function which transforms non-linear input data to a linear one in a high-dimensional feature space. Kernel function is generally a non-linear function. Widely used kernel functions are radial basis function (RBF) kernel and polynomial kernel.

The parameters are estimated using $\epsilon$-insensitivity loss function[2], which is defined as:

$$L_\epsilon(f(x)-y)=\begin{cases}|f(x)-y|-\epsilon & if \quad |f(x)-y|\geq\epsilon \\ 0 & otherwise\end{cases} \tag{9}$$

where $y$ is the desired(target) output; $\epsilon$ is defined as the region of $\epsilon$-insensitivity.

SVR works on the principle of structural risk minimization. When empirical risk [3]and structure risk are considered together, the objective function is defined as:

$$Min: \frac{1}{2}\mathbf{z}^T\mathbf{z}+C\sum_i(\xi_i+\xi_i^*)$$
$$subject \quad to \begin{cases} y_i-\mathbf{z}^T x_i-b\leq\epsilon+\xi_i \\ \mathbf{z}^T x_i+b-y_i\leq\epsilon+\xi_i^* \\ \xi_i,\xi_i^*\geq 0\end{cases} \tag{10}$$

where, $i$ = *1,..., n* is the number of training data; $(\xi_i+\xi_i^*)$ is the empirical risk; $\frac{1}{2}\mathbf{z}^T\mathbf{z}$ is the structure risk preventing over-learning and lack of applied universality; and $C$ is a modifying coefficient representing the trade-off between empirical risk and structure risk.

The optimal value of weights can be solved using Lagrange condition for various values of modifying coefficient $C$, band area width $\epsilon$, and kernel function $K$. The approximation function of the SVR-based regression is defined as

$$f(x,\mathbf{z})=f(x,\alpha,\alpha^*)=\sum_{i=1}^{N}(\alpha_i-\alpha_i^*)K(x,x_i)+b \tag{11}$$

where $\alpha_i$ and $\alpha_i^*$ are Lagrangian multipliers which satisfy the equality condition $\alpha_i\alpha_i^*=0$.

A survey on use of SVR for time series prediction is carried out by Sapankevych & Sankar (2009)[22].

[2] Other loss function is quadratic loss function but limitation is that it depends on outliers
[3] Empirical risk represents the error generated by the estimation process of the value

### 2.4 Proposed Hybrid Models

The proposed hybrid models (MODWT-ANN and MODWT-SVR) for prediction of stock price integrate the advantages of a decomposition model (namely, MODWT) and machine learning models (namely, ANN and SVR). The steps (refer Figure 5) are enumerated below:

1. The original series is decomposed into a set of various sub-series (also known as wavelets) using DWT.
2. Each sub-series is forecasted separately using ANN and SVR in case of the hybrid MODWT-ANN and MODWT-SVR models, respectively.
3. Forecasted sub-series are recombined to obtain the final forecast value.
4. Performance/error measures are calculated by comparing the final forecast value (obtained in the previous step) with the original series.
5. To understand the effectiveness of the proposed models, Wilcoxon-Signed Rank Test, a non-parametric statistical model is used to compare the predictive capability of MODWT-ANN and MODWT-SVR with ANN and SVR models (without decomposition).

## 3. Empirical Analysis

### 3.1 Data

National Stock Exchange Fifty (commonly known as NIFTY 50 or NIFTY) and Bombay Stock Exchange Sensitive Index (commonly known as SENSEX) are the prominent stock indices for Indian equity market.

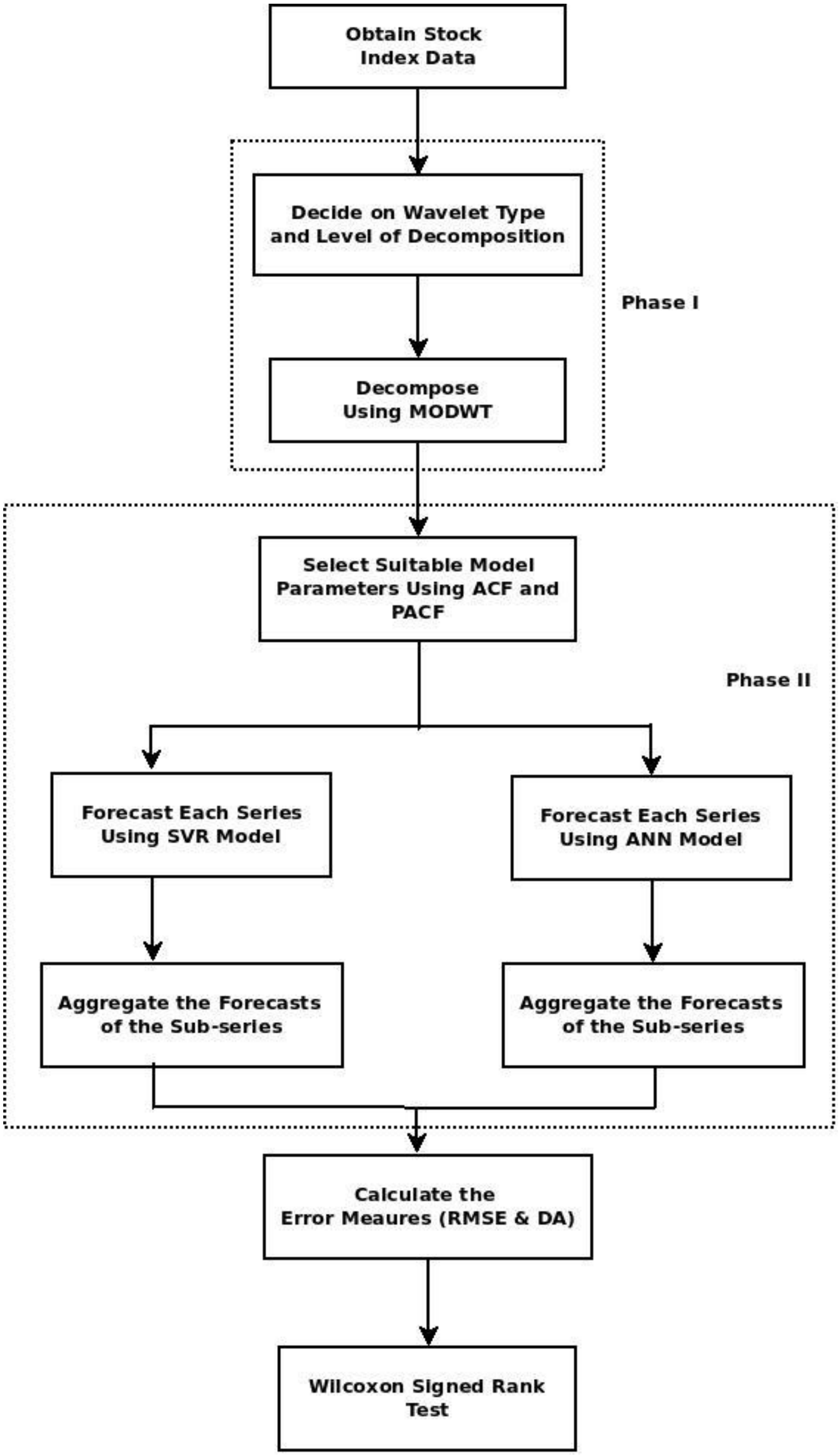

**Figure 5: Flow Chart of the Proposed Hybrid Models**

Founded in 1995, National Stock Exchange Fifty index is the benchmark index of National Stock Exchange of India and Bombay Stock Exchange Sensitive Index is the benchmark index of Bombay Stock Exchange. The former consists of 50 stocks representing 22 sectors while the latter comprises of only 30 stocks. National Stock Exchange Fifty index is a better representative of Indian stock market.

The index considered for this study is National Stock Exchange Fifty index. The data sample comprised of weekly closing prices covering a period of 8 years ranging from September 2007 to July 2015, thus, resulting in total of 409 data points. The period considered for the study covers the 2008 Financial Crisis. The world economy including the Indian economy was affected. The reasons for decline of the Indian stock market are quoted as: (i) difficulties were experienced while raising funds in the domestic market, (ii) Foreign Institutional Investor (FII) funds were withdrawn from the stock markets, (iii) credit crunch in banks. The performance of National Stock Exchange Fifty index is shown in Figure 6.

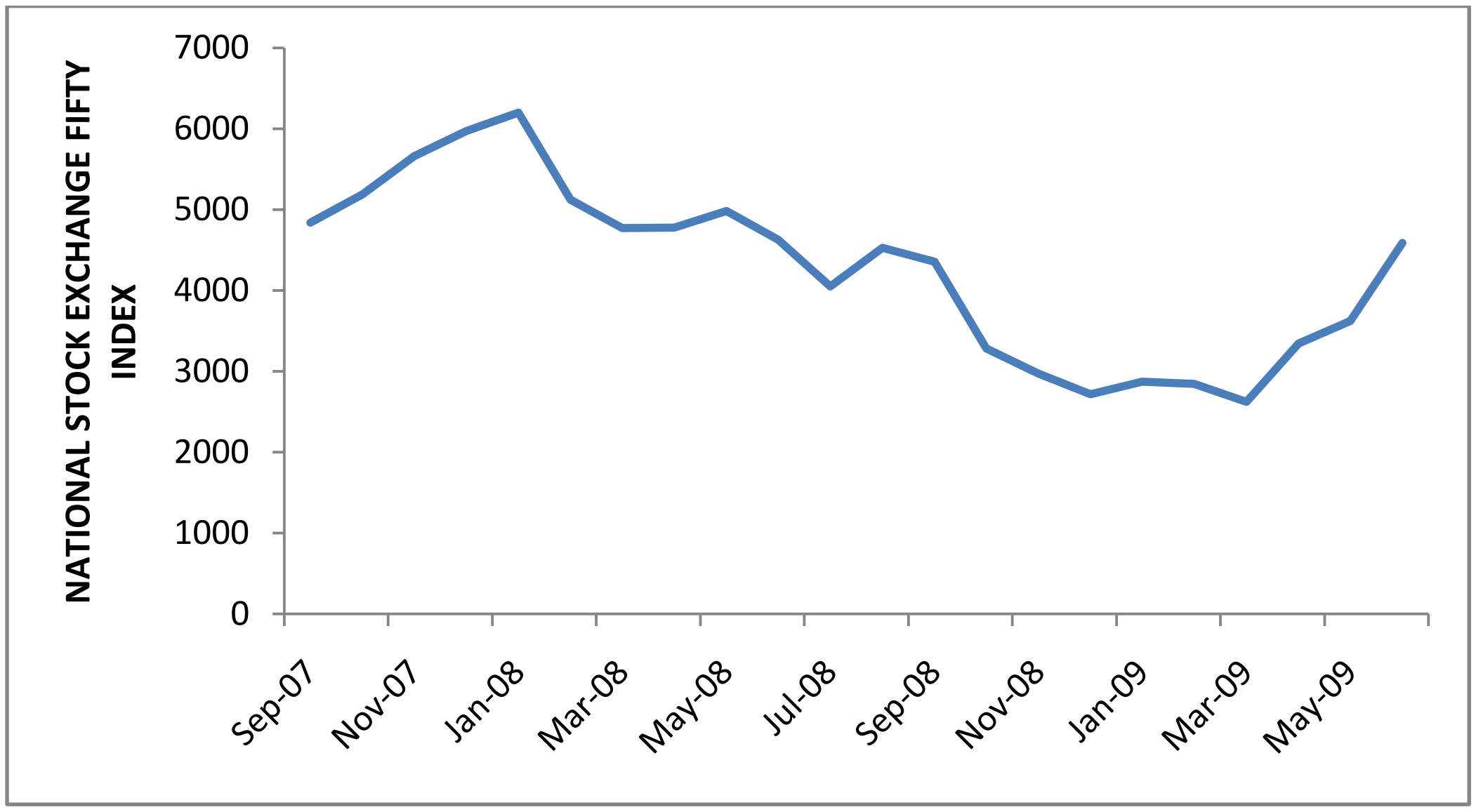

**Figure 6: Performance of National Stock Exchange Fifty Index during 2008 Financial Crisis**

### 3.2 Phase I: Decomposition of National Stock Exchange Fifty index

a. **Type of Wavelet Filter and Level of Decomposition:** MODWT is used to decompose the weekly National Stock Exchange Fifty index closing price to overcome the limitation of DWT (refer to Section 2.1). The type of wavelet filter chosen here is "Haar". Haar wavelet helps to capture fluctuations between adjacent observations and is known as differencing filter. It also eliminates the problem of aliasing. The level of decomposition to be selected depends on the number of data points available. According to Tsai & Chiang (2003)[23], all the hidden information in the data is revealed when the level of decomposition is 6. However, when the level of decomposition increases, the data will be smoothened to a larger extent, thus, leading to loss of information (Lahmiri, 2014). Hence, the level of decomposition chosen here is 3.

b. **Decomposition:** MODWT (type: Haar) was used to decompose the National Stock Exchange Fifty index into various subseries also known as wavelets. Here, we obtained three wavelet co-efficients ($W_1$, $W_2$, and $W_3$) and one scaling co-efficient ($V_3$). The MODWT decomposition of the series is shown in Figure 7. The first row represents the scaling co-efficient $V_3$. This is the smoothed component whose pattern is quite similar to the original series. This represents the trend of the movement of the stock index. Rows 2, 3 and 4 represent the wavelet coefficients $W_1$, $W_2$ and $W_3$, respectively. $W_1$ is the high frequency component. The last row represents the original time series $x(t)$.

### 3.3 Phase II: Prediction of Subseries

In this phase, each subseries obtained is predicted using two machine learning models, namely, ANN and SVR. The parameters for both the models are selected initially using Autocorrelation Function (ACF) and Partial Autocorrelation Function (PACF).

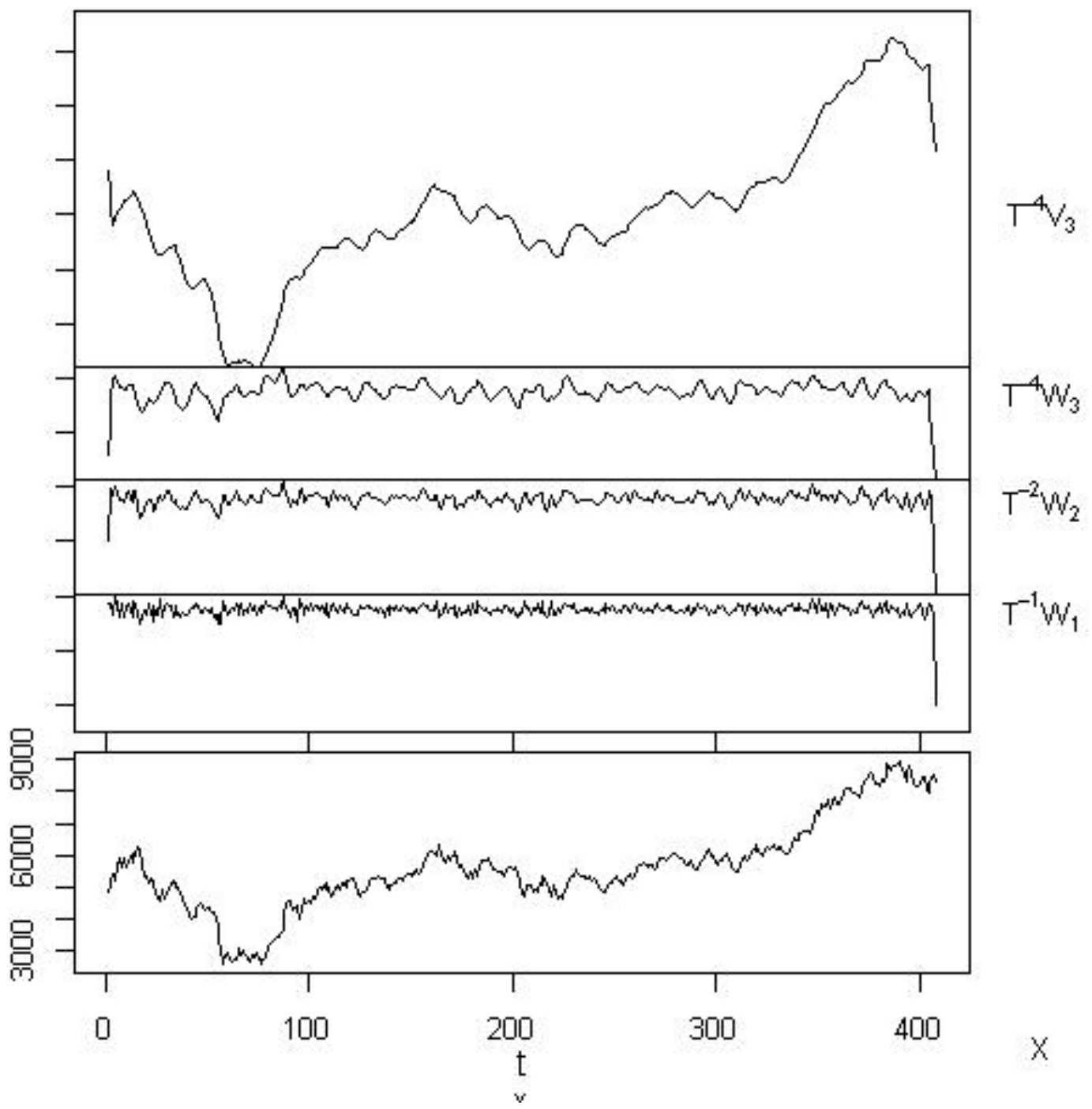

**Figure 7: Wavelets of National Stock Exchange Fifty Index**

a. **Check for Stationarity:** In this step, each subseries obtained is checked for stationarity using the widely used Augmented Dickey-Fuller (ADF) test (Dickey & Fuller, 1979[24]; Dickey & Fuller, 1981[25]). Here, the subseries $W_1$ and $W_2$ were found to be stationary, while $W_3$ and $V_3$ were not. If a series is found to be non-stationary, then first difference [4]of the series is obtained and checked for stationarity using ADF test. The process is continued till the series becomes stationary or on reaching maximum number of iterations. The first difference of $W_3$ and $W_4$ were found to be stationary.

b. **Identification of Lag Parameter:** The relationship between a series and its past values, which is estimated as the lag parameter, is used as the input to the ANN and SVR in their respective hybrid models. ACF and PACF are used to determine the lags in each sub-series.

[4] difference of the first difference of the series. Suppose *F(t) = y(t), y(t-1)... y(t-n)*, then the first difference is *d1 = {y(t-1)-y(t)}, {y(t-2)-y(t-1)},...* and the second difference *d2 = {y(t-2)-2y(t-1)+y(t)}, {y(t-3)-2y(t-2)+y(t+1)},...*

The ACF and PACF plots of $W_2$ are shown in Figure 8. The ACF plot shows a trailing pattern and PACF plot displays a decreasing and oscillating pattern. This indicates the existence of AR component. From both plots, it can be seen that the sub-series $W_2$ cuts off at lag 4. So, the value of $W_2$ at time $t$ is dependent on its past 4 values. The data format of the same can be expressed as:

$$X(t) = f\left(X(t-1), X(t-2), X(t-3), X(t-4)\right) \tag{12}$$

In the similar way, the lag parameters of other sub-series are determined.

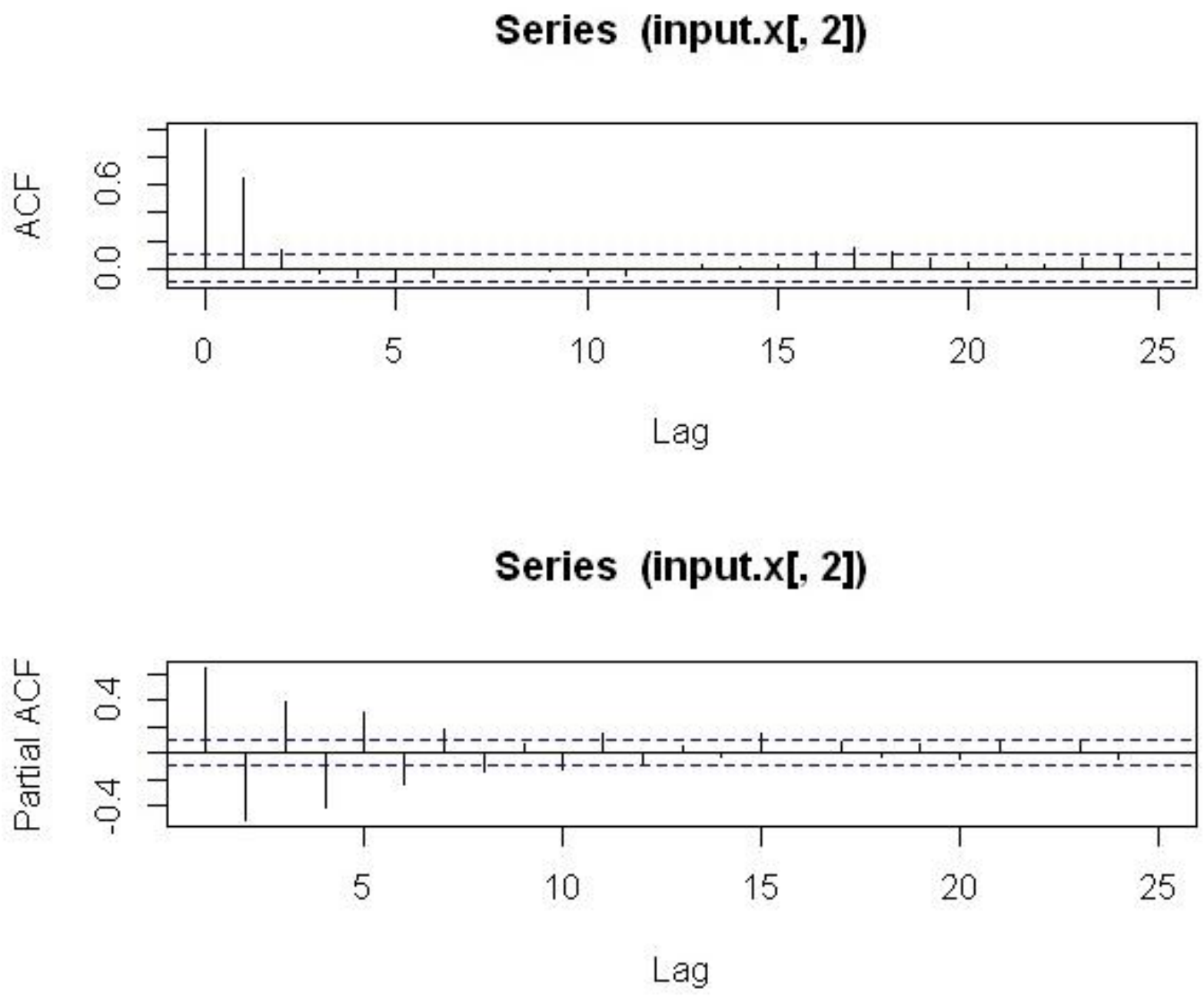

**Figure 8: ACF and PACF of wavelet coefficient $W_2$**

c. **Prediction:** 1-step ahead forecasts for each series were obtained using ANN (in case of hybrid MODWT-ANN model) and SVR (in case of hybrid MODWT- SVR model). 70% of the data was used for training the model and remaining 30% for testing the model.

As a data preprocessing step for both ANN and SVR models, the data is normalized between [-1,1], [0,1] and using z-scores. It helps the models to overcome the limitations of getting trapped in local minima and overfitting. Further, it quickens the training process (Crone, Guajardo, & Weber, 2006[26]; Wu & Lo, 2010[27]). The model using z-score as a preprocessing step seemed to perform better than remaining two normalization processes.

- **ANN Model:** As mentioned earlier, the neural network model has three layers and each layer consisting of certain number of neurons. Since it is a prediction problem, the number of neurons in output layer is 1. Number of neurons in the input layer is obtained using lag parameter. For instance, in case of $W_1$, the lag parameter identified was 4, hence the number of neurons in the input layer is 4. This is mathematically expressed as:

$$W_1(t) = f\left[W_1(t-1), W_1(t-2), W_1(t-3), W_1(t-4)\right] \tag{13}$$

There are no defined rules to identify the number of neurons in the hidden layer, hence, they are selected on basis of best performances of the model. The number of neurons in each layer is presented in Table 1.

The model is trained using Resilient Propagation (RP) algorithm (Riedmiller & Braun, 1993[19]). RP is preferred over the most commonly used BP algorithm because of its superior performance (Liu et al., 2012[13]). Further, the training of neural network model is faster and does not require specification of parameters such as learning rate and momentum during the training process.

**Table 1: Number of Neurons in Various Layers of ANN**

| Wavelets | Number of Neurons | | |
|---|---|---|---|
| | Input Layer | Hidden Layer | Output Layer |
| $W_1$ | 4 | 10 | 1 |
| $W_2$ | 4 | 10 | 1 |
| $W_3$ | 7 | 10 | 1 |
| $V_3$ | 5 | 10 | 1 |

- **SVR Model:** Here, similar to ANN model, the wavelets and their previous values (identified based on lag parameter) are used as input variables. Since SVR models are sensitive to the parameters selected ($C$ and $\epsilon$), grid search proposed by Lin, Hsu, & Chang (2003)[28] is adopted in this paper. In this approach, the best values of $\epsilon$ and $C$ are determined by exponentially growing sequence of $\epsilon$ and $C$ ($C = 2^{-3}, 2^{-1}, \ldots, 2^{15}$). Best parameters help to minimize the forecasting mean square error. Analysis is carried out in R language, a statistical computing language and environment.

Before recombining the forecast of the subseries, all the transformations applied to the data should be removed. For instance, the data should be denormalized. In case of first difference and second difference, the transformations should be reverted to the original form.

d. **Obtaining the Final Forecasts:** The forecasts of subseries obtained using ANNs are summed up to obtain the final forecast. Similarly, the final forecast is obtained for the ones generated using SVR model. The final forecasts can be obtained by using the following equation:

$$X'(t) = W'_1(t) + W'_2(t) + W'_3(t) + V'_3(t) \tag{14}$$

where, $X^{'}(t)$ is the final forecast

$W'_1(t)$, $W'_2(t)$, $W'_3(t)$ and $V'_3(t)$ are the respective predicted values of $W_1(t)$, $W_2(t)$, $W_3(t)$ and $W_3(t)$ obtained using ANN and SVR models.

The final forecasts obtained are then compared with the original series to calculate error measures.

To analyze the effectiveness of the proposed approach (hybrid MODWT-ANN and MODWT-SVR models), the original series (i.e. series without MODWT decomposition) is modeled and predicted using SVR and ANN models.

## 4. Results and Discussion

### 4.1 Error Measures

The forecasting performances of various predictive models, namely, ANN, SVR, MODWT-ANN and MODWT-SVR are evaluated based on two performance measures: Root Mean Square Error (RMSE) and Directional Accuracy (DA).

The difference between the original series ($X_i(t)$) and the corresponding predicted values ($F_i(t)$) is known as Error ($E_i(t)$). Root Mean Square Error is calculated as the square root of mean of error values. Lower is the value of RMSE, better the predictive model. RMSE is expressed mathematically as:

$$RMSE = \sqrt{\sum_{i=1}^{n} E_i(t)^2 / n} \tag{15}$$

where, $E_i(t) = X_i(t) - F_i(t)$ (16)

Directional Accuracy, expressed in percentage, represents the number of times the predicted values matched the direction of the sign followed by original series. Higher the value of DA, better are the forecasts. These error measures are calculated using the predicted values of test dataset.

The performance measures of the four forecast models are represented in Table 2. From the table, it can be seen that the decomposition-based hybrid models have better performances compared to those without decomposition. The hybrid MODWT-SVR model has a remarkably better performance, in terms of both RMSE and DA, in comparison to other models. Hybrid MODWT-SVR model has the lowest RMSE of 50.08 and a directional accuracy of 80%. The model has accurately predicted the direction of movement of the stock prices 80% of the times. With RMSE and DA of 155.45 and 74.95%, respectively, the hybrid MODWT-ANN model performed considerably better than both ANN and SVR models. The results of the 1-step ahead forecasts obtained using all the four models are represented in Figure 9.

**Table 2: Error Measures**

| Model | RMSE | DA (%) |
|---|---|---|
| ANN | 184.28 | 53.06 |
| SVR | 169.45 | 58.16 |
| Hybrid MODWT-ANN | 155.45 | 74.95 |
| Hybrid MODWT-SVR | 50.08 | 80.00 |

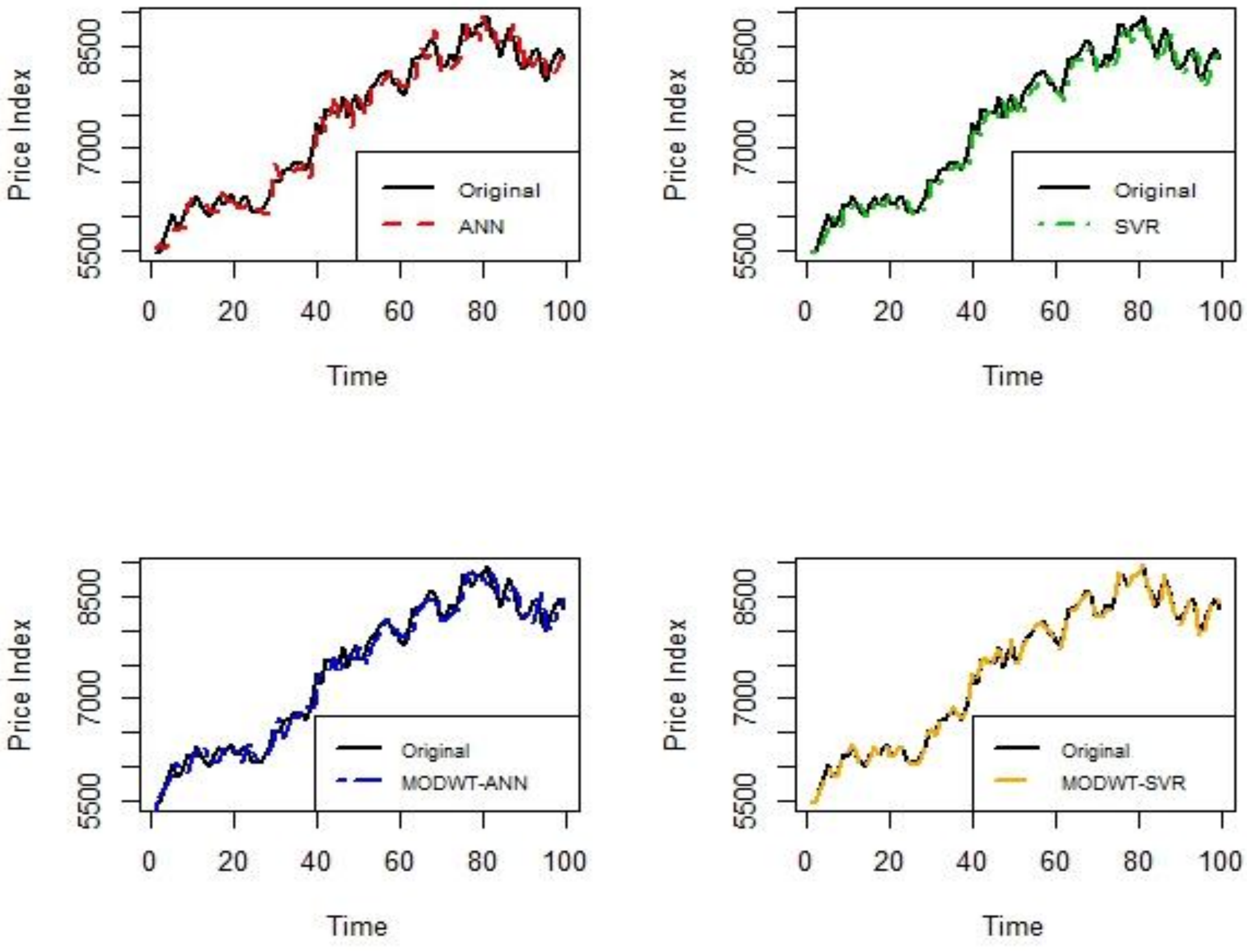

**Figure 9: Comparison of 1-step ahead forecasts of ANN, SVR, hybrid MODWT-ANN model and hybrid MODWT-SVR model.**

### 4.2. Significance Test

The predictive capabilities of two different models are evaluated using a popular non-parametric and distribution-free technique known as Wilcoxon Signed-Rank Test (WSRT) (Diebold & Mariano, 1995[29]; Kao, Chiu, Lu, & Chang, 2013[30]). In this test, the signs and the ranks of the predicted values are compared to identify whether two predictive models are different. Hence, WSRT is used to evaluate whether performance of the presented hybrid MODWT-SVR and MODWT-ANN models is better than that of SVR and ANN models without decomposition.

Two-tailed WSRT was carried out to evaluate the predictive performance of the proposed models (hybrid MODWT-ANN and hybrid MODWT-SVR) and remaining models (ANN and SVR models). Results of the test are shown in Table 3. From the table, it can be seen that z statistics value is beyond (-1.96, 1.96), hence the null hypothesis of two models being same is not accepted. The results are significant at 99% confidence level ($\alpha = 0.01$). The sign "+" in the table represents that predictive capability of hybrid MODWT-SVR model is superior than two traditional computational intelligence techniques, namely, SVR and ANN. Signs "-" and "=" refer to underperformance and similar performance of the hybrid models compared to other two models, respectively. The WSRT results confirm that the hybrid MODWT-SVR and MODWT-ANN models outperformed the traditional SVR and ANN models.

**Table 3: WSRT between hybrid MODWT-ANN, hybrid MODWT-SVR, ANN and SVR models**

| | **Hybrid MODWT-ANN** | | **Hybrid MODWT-SVR** | |
|---|---|---|---|---|
| | **z** | **WSRT** | **z** | **WSRT** |
| ANN | -3.188 | + | -3.980 | + |
| SVR | -4.509 | + | -4.558 | + |

+ : DWT-SVR > SVR, DWT-SVR > ANN, DWT-ANN > SVR, DWT-ANN > ANN
= : DWT-SVR = SVR, DWT-SVR = ANN, DWT-ANN = SVR, DWT-ANN = ANN
- : DWT-SVR < SVR, DWT-SVR < ANN, DWT-ANN < SVR, DWT-ANN < ANN

### 4.3. Buying and Selling Decisions

Using the forecasting models discussed above the closing price of a stock on the first trading day of the following week can be predicted with reasonable accuracy. These predictions can be used by the investors to make investment related decisions using few trading rules. The predictions of MODWT-SVR model are used because of high directional accuracy and low RMSE values.

Let $\hat{y}_k$ and $y_k$ be the forecasted closing price and actual closing price respectively on first trading day in the $k^{th}$ trading week. Consider a situation where the closing price is expected to rise in the trading week $k$ but it falls or remains same, then it is represented using an error index $E_k$ (Hsu, 2014[31]), which is defined as follows:

$$E_K = \begin{cases} 1 \quad if \quad \hat{y}_k > y_{k-1} \quad and \quad y_k \le y_{k-1} \quad \forall \quad k = 2,3,...,n \\ 0 \quad otherwise \end{cases} \tag{17}$$

where, $n$ is the total number of weeks.

Though the analysis has been carried out using the index values, the same model and trading rules can be used for stocks as well. Based on the error index ($E_k$), following three rules [5]are used in this study:

**Rule 1:** IF ( $\hat{y}_{k+1} > y_k$ ) AND (if an investor is not holding any stock on first trading day in the $k^{th}$ week) THEN (he/she is advised to buy the stock on the next trading day in the $k^{th}$ week)

**Rule 2:** IF ($\hat{y}_{k+1} < y_k$) AND (if an investor is holding any stock on first trading day in the $k^{th}$ week) THEN (he/she is advised to sell the stock on the next trading day in the $k^{th}$ week)

**Rule 3:** IF (the investor is holding any stock on first trading day in the $k^{th}$ trading week) AND ( $\sum_{j=0}^{2} E_{k-j} = 3$) THEN (the investor is advised to sell the stock on the second trading day in $k^{th}$ trading week)

[5] Rules have been explained with respect to stocks though the analysis and illustration have been carried out using index values

The first rule suggests that an investor should buy a stock if he does not hold any stock on first day of the trading week since the price in the following week is expected to rise. The second rule suggests that if an investor is holding a stock on the first day of the trading week and observes that price is expected to fall the next week, he is advised to sell his stocks. The final rule suggests the investor to sell his held stock if the predictions of rising stock price are completely wrong for three consecutive weeks.

The return on investment is calculated using the following equation:

$$ROI = \prod_{i=1}^{k} \frac{C_{S_i} - C_{B_i}}{C_{B_i}} \qquad (18)$$

where, $C_{Si}$ represents the selling price of the $i^{th}$ transaction, $C_{Bi}$ represents the buying price of the $i^{th}$ transaction and $k$ is the total number of transaction.

Here, it is assumed that the security can be sold and bought at the opening price on the next trading day of the week once buying/selling decisions are taken. The transactions based on the trading rules are illustrated in Table 4.

**Table 4: Illustration of Implementation of Trading Rules**

| Date | Closing Value | Forecasted Closing Value[6] | Transaction | Transaction Date | $E_k$ | Rule |
|---|---|---|---|---|---|---|
| 5-Jan-15 | 8284.50 | 8392.73 | Buy at 8325.30 | 6-Jan-15 | 1 | Rule 1 |
| 12-Jan-15 | 8513.80 | 8852.33 | - | - | 0 | |
| 19-Jan-15 | 8835.60 | 8743.47 | Sell at 8575.09 | 20-Jan-15 | 0 | Rule 2 |
| 27-Jan-15 | 8808.90 | 8643.77 | - | - | 0 | |
| 2-Feb-15 | 8661.05 | 8803.13 | Buy at 8823.15 | 3-Feb-15 | 1 | Rule 1 |
| 9-Feb-15 | 8805.50 | 8837.31 | - | - | 0 | |
| 16-Feb-15 | 8833.60 | 8834.68 | - | - | 0 | |

[6] The forecasted closing price for the first trading day in the next trading week

| | | | | | | |
|---|---|---|---|---|---|---|
| 23-Feb-15 | 8844.60 | 8974.32 | - | - | 0 | |
| 2-Mar-15 | 8937.75 | 8669.78 | Sell at 8962.85 | 3-Mar-15 | 0 | Rule 2 |
| 9-Mar-15 | 8647.75 | 8462.15 | - | - | 0 | |
| 16-Mar-15 | 8570.90 | 8341.93 | - | - | 0 | |
| 23-Mar-15 | 8341.40 | 8433.20 | Buy at 8537.05 | 24-Mar-15 | 1 | Rule 1 |
| 30-Mar-15 | 8586.25 | 8833.33 | - | - | 0 | |
| 6-Apr-15 | 8780.35 | 8546.42 | Sell at 8684.45 | 7-Apr-15 | 0 | Rule 2 |
| 13-Apr-15 | 8606.00 | 8283.79 | - | - | 0 | |
| 20-Apr-15 | 8305.25 | 8087.78 | - | - | 0 | |
| 27-Apr-15 | 8181.50 | 8086.90 | - | - | 0 | |
| 4-May-15 | 8191.50 | 8231.53 | Buy at 8338.40 | 5-May-15 | 0 | Rule 1 |
| 11-May-15 | 8262.35 | 8457.66 | - | - | 0 | |
| 18-May-15 | 8458.95 | 8472.62 | - | - | 0 | |
| 25-May-15 | 8433.65 | 8048.21 | Sell at 8377.10 | 26-May-15 | 0 | Rule 2 |
| 1-Jun-15 | 8114.70 | 7888.53 | - | - | 0 | |
| 8-Jun-15 | 7982.90 | 8149.56 | Buy at 8026.50 | 9-Jun-15 | 1 | Rule 1 |
| 15-Jun-15 | 8224.95 | 8371.96 | - | - | 0 | |
| 22-Jun-15 | 8381.10 | 8501.96 | - | - | 0 | |
| 29-Jun-15 | 8484.90 | 8417.48 | Sell at 8316.35 | 30-Jun-15 | 0 | Rule 2 |

The closing value of the index on January 5, 2015 is 8284.5, which is less than the forecasted value of first trading day of next week (8392.73). Hence, based on Rule 1, the investor is advised to buy the security on January 6, 2015. On January 19, 2015, since the forecasted value for next week (January 27, 2015) is lower than the current value, then the investor is advised to sell his security on January 20, 2015. Similarly, on February 2, 2015, the stock price (8661.05) is less than that of the forecasted value for February 9, 2015 (8803.127), hence the investor is advised to buy security on February 3, 2015 at the opening price. On February 9, 2015, the closing value is less than that of forecasted value, hence no action is taken. The stock is held till March 2, 2015 when the forecasted value (8669.78) of first trading day of the subsequent week (i.e., March 9, 2015) is lower than the current vale (8937.75). Using Rule 2, the investor is advised to sell the security. The trading rules were applied to the test data set. It was found that return on investment obtained based on trading rules using predictions of MODWT-SVR and trading rules was higher than that of Buy-and-Hold strategy.

## 5. Conclusion

The paper proposed and presented a hybrid forecasting approach integrating the advantages of a decomposition model and machine learning models. Two hybrid models, namely, MODWT-ANN and MODWT-SVR are presented to predict 1-step ahead forecasts of National Stock Exchange Fifty index. The hybrid approach first used MODWT to decompose the time series data. Then, it uses SVR and ANN in their respective models to predict each subseries independently and then the forecasted sub-series are aggregated to obtain the final forecasts. The presented models showed a consistent superior performance in predicting the weekly National Stock Exchange Fifty index, as compared to both ANN and SVR, which indicate their ability to forecast non-stationary and non-linear time series. From the experiments, it can be observed that decomposition of the series using MODWT assisted in improving the performance of the machine learning models quite significantly. The performance measures obtained reiterate the fact the forecasts obtained by aggregating the predicted components of a time series are more accurate than the forecasts obtained by modeling the original series itself.

The proposed model is advantageous for predicting the stock index since it is a decomposition-based model. It helps to deconstruct the time-series into various frequency components. Individual components are easier to predict than original non-stationary and non-linear data series. The second part of the model uses SVR which has robust performance than ANN model. Since the proposed model is data adaptive model, the trend in the time series is captured and this helps in increasing the accuracy of the prediction.

Further, two trading strategies, namely, based on MODWT-SVR predictions and Buy-and-hold strategies were evaluated to determine the timing for buying and selling the securities. It was found that the trading strategies based on the results of MODWT-SVT yielded better ROI than that of Buy-and-Hold strategy.

Here, in this paper, the analysis is carried out using “haar” filter and the level of decomposition assumed is three. The work can be extended to other levels to study the effect of other filters, namely, db4, db6, db8 and db1 and on various levels of decomposition can also be carried out. As a part of future direction, the model can be tested for high frequency intraday stock index or stock price data. In order to capture the effects of macro-economic variables on stock price, multivariate forecasting models can be used for predicting the stock price.

## References

1. Novak, M. G., & Velušček, D., Prediction of stock price movement based on daily high prices. **Quantitative Finance**, doi: 10.1080/14697688.2015.1070960 (2015).
2. Fama, E. F., Efficient capital markets: A review of theory and empirical work. **Journal of Finance** (25: 1970)
3. Lo, A. W., & MacKinlay, A. C., Stock market prices do not follow random walks: Evidence from a simple specification test. **Review of Financial Studies** (1: 1988).
4. Chen, N.-F., Financial investment opportunities and the macroeconomy. **Journal of Finance** (46:1991).
5. Bilson, C., Brailsford, T., & Hooper, V. J., Selecting macroeconomic variables as explanatory factors of emerging stock market returns. **Pacific Basin Finance Journal** (9: 2001).
6. Yao, J., Tan, C., & Poh, H.-L., Neural networks for technical analysis: A study on KLCI. **International Journal of Theoretical and Applied Finance** (2: 1999).
7. Bettman, J. L., Sault, S., & Schultz, E., Fundamental and technical analysis: Substitutes or complements? **Accounting & Finance** (49: 2009).

8. Atsalakis, G., & Valavanis, K., Surveying stock market forecasting techniques- Part I: Conventional methods. In C. Zopounidis (Ed.), **Computation optimization in economics and finance research compendium** (p. 49- 104). New York: Nova Science Publishers, Inc. (2013).
9. Atsalakis, G., & Valavanis, K., Surveying stock market forecasting techniques- Part II: Soft computing methods. **Expert Systems with Applications** (36: 2009).
10. Bollerslev, T., Generalized autoregressive conditional heteroskedasticity. **Journal of Econometrics** (31: 1986).
11. Matei, M., Assessing volatility forecasting models: Why GARCH models take the lead. **Journal for Economic Forecasting** (4: 2009).
12. Theodosiou, M., Forecasting monthly and quarterly time series using STL decomposition. **International Journal of Forecasting** (27: 2011).
13. Liu, H., Chen, C., Tian, H., & Li, Y., A hybrid model for wind speed prediction using empirical mode decomposition and artificial neural networks. **Renewable Energy** (48: 2012).
14. Lahmiri, S., Wavelet low- and high-frequency components as features for predicting stock prices with backpropagation neural networks. **Journal of King Saud University - Computer and Information Sciences**, (26: 2014).
15. Ortega, L., & Khashanah, K., A neuro-wavelet model for the short-term forecasting of high-frequency time series of stock returns. **Journal of Forecasting** (33: 2014).
16. Murtagh, F., Starck, J., & Renaud, O., On neuro-wavelet modeling. **Decision Support Systems** (37:2004)
17. Li, T., Li, Q., Zhu, S., & Ogihara, M. A survey on wavelet applications in data mining. **SIGKDD Explorations Newsletter** (4: 2002).
18. Gençay, R., Selçuk, F., & Whitcher, B., Discrete wavelet transforms. In R. G. S. Whitcher (Ed.), **An introduction to wavelets and other filtering methods in Finance and economics** (p. 96 - 160). San Diego: Academic Press (2002).
19. Riedmiller, M., & Braun, H., A direct adaptive method for faster back-propagation learning: The RPROP algorithm. In **IEEE international conference on neural networks** (Vol. 1: 1993).

20. Al-Hnaity, B., & Abbod, M., A novel hybrid ensemble model to predict FTSE100 index by combining neural network and EEMD. In **2015 European Control Conference (ECC)** (p. 3021-3028: 2015).
21. Vapnik, V. N., **The nature of statistical learning theory**. New York, NY, USA: Springer-Verlag New York, Inc. (1995)
22. Sapankevych, N., & Sankar, R., Time series prediction using support vector machines: A survey. **Computational Intelligence Magazine**, IEEE (4: 2009).
23. Tsai, D.-M., & Chiang, C.-H., Automatic band selection for wavelet reconstruction in the application of defect detection. **Image and Vision Computing** (21: 2003).
24. Dickey, D. A., & Fuller, W. A., Distribution of the estimators for autoregressive time series with a unit root. **Journal of the American Statistical Association** (74: 1979).
25. Dickey, D. A., & Fuller, W. A., Likelihood ratio statistics for autoregressive time series with a unit root. **Econometrica** (49: 1981).
26. Crone, S., Guajardo, J., & Weber, R., The impact of preprocessing on support vector regression and neural networks in time series prediction. In **Proceedings of the International Conference on Data Mining (DMIN '06)** (p. 37-42). Las Vegas, USA,: CSREA (2006).
27. Wu, G., & Lo, S., Effects of data normalization and inherent-factor on decision of optimal coagulant dosage in water treatment by artificial neural network. **Expert Systems with Applications** (37: 2010).
28. Lin, C., Hsu, C., & Chang, C., **A practical guide to support vector classification** (Tech. Rep.). National Taiwan University, Taipei: Department of Computer Science and Information Engineering (2003).
29. Diebold, F. X., & Mariano, R. S., Comparing predictive accuracy. **Journal of Business and Economic Statistics** (13: 1995).
30. Kao, L.-J., Chiu, C.-C., Lu, C.-J., & Chang, C.-H., A hybrid approach by integrating wavelet-based feature extraction with MARS and SVR for stock index forecasting. **Decision Support Systems** (54: 2013).
31. Hsu, C. –M., An integrated portfolio optimisation procedure based on data envelopment analysis, artificial bee colony algorithm and genetic programming. **International Journal of Systems Science**, (45: 2014).

## Biography

**Dhanya Jothimani** is a Ph.D. scholar in the Department of Management Studies at IIT Delhi, India. She received her Bachelors of Technology (B.Tech.) from National Institute of Technology, Tiruchirappalli, India, and Masters of Technology (M.Tech.) from IIT Kharagpur, India. Her research interests include Financial Analytics, Data Mining, and Operations Research. She is a member of Institute for Operations Research and Management Science (INFORMS), American Finance Association (AFA), Decision Science Institute (DSI) and Association of Information Systems (AIS).

**Ravi Shankar** is a Professor of Supply Chain and Operations Management at IIT, Delhi, India. His areas of interest are supply chain management, business analytics, operations research, fuzzy modeling, six-sigma, project management, strategic technology management, etc. He has published over 300 research papers in journals and conferences, including Omega, European Journal of Operations Research, International Journal of Production Research, International Journal of Production Economics, International Journal of Supply Chain Management, International Journal of Physical Distribution & Logistics Management, IEEE Systems Man and Cybernetics Part C, Computers and Industrial Engineering, Computers and Operations Research, etc.

**Surendra S. Yadav** received his Bachelor of Technology (B. Tech.) from IIT, Kanpur, India, MBA from University of Delhi, India, and PhD in Management from University of Paris 1 Pantheon-Sorbonne, France. He is a Professor in the Department of Management Studies, IIT, Delhi, India. He teaches Corporate Finance, International Finance, International Business and Security Analysis & Portfolio management. His research interests are in all these areas and general management. He is the Editor-in-Chief of Journal of Advances in Management Research, published by Emerald, UK.